\documentclass[aps,prl,reprint,twocolumn,amsmath,amssymb,citeautoscript,longbibliography]{revtex4-1}
\usepackage[colorlinks=true, linkcolor=blue, citecolor=blue, urlcolor=blue]{hyperref}
\usepackage{hyperref}

\usepackage[utf8]{inputenc} 
\DeclareUnicodeCharacter{0131}{\i} 
\DeclareUnicodeCharacter{0301}{\'{}} 

\usepackage{graphicx}
\usepackage{appendix}
\usepackage{csquotes}
\usepackage{dcolumn}
\usepackage{bm}
\usepackage{latexsym,epsfig}
\usepackage{graphicx}
\usepackage{verbatim}
\usepackage{comment}
\usepackage{amsmath}
\usepackage{amssymb}
\usepackage{physics}
\usepackage{stmaryrd}
\usepackage{color}
\usepackage{epstopdf}
\usepackage{grffile}
\usepackage{lipsum}
\usepackage{enumitem}
\usepackage{ulem}

\usepackage[dvipsnames]{xcolor}
\DeclareGraphicsExtensions{.eps}

\newcommand{\beq}{\begin{equation}}
\newcommand{\eeq}{\end{equation}}
\newcommand{\bea}{\begin{eqnarray}}
\newcommand{\eea}{\end{eqnarray}}
\newcommand{\ben}{\begin{eqnarray*}}
\newcommand{\een}{\end{eqnarray*}}
\newcommand{\bfig}{\begin{figure}}
\newcommand{\efig}{\end{figure}}

\usepackage{xcolor}  
\definecolor{RED}{RGB}{220,0,0}

\usepackage{hyperref}

\usepackage{hyperref}
\hypersetup{
    colorlinks=true,      
    urlcolor=blue,
    citecolor=blue,
    linkcolor=blue
}

\begin{document}

\title{Multiple Many-Body Localization Transitions in Driven Non-Hermitian Chains with Spatially Correlated Potentials}

\author{Sanchayan Banerjee$^{1,2,\# }$\thanks{These authors contributed equally to this work.}}
\email{sanchayan.banerjee@niser.ac.in}
\author{Ayan Banerjee$^{3,\# }$\thanks{These authors contributed equally to this work.}}
\email{ayan.banerjee@mpl.mpg.de}
\author{Tapan Mishra$^{1,2}$}
\email{mishratapan@niser.ac.in}
\author{Flore K. Kunst$^{3,4}$}
\email{flore.kunst@mpl.mpg.de}

\thanks{$^\#$These authors contributed equally to this work.}

\affiliation{$^1$School of Physical Sciences, National Institute of Science Education and Research, Jatni,  Odisha 752050, India}
\affiliation{$^2$Homi Bhabha National Institute, Training School Complex, Anushaktinagar, Mumbai, Maharashtra 400094, India}
\affiliation{$^3$Max Planck Institute for the Science of Light, 91058 Erlangen, Germany}
\affiliation{$^4$Department of Physics, Friedrich-Alexander-Universit\"at Erlangen-N\"urnberg, 91058 Erlangen, Germany}

\date{\today}

\begin{abstract}

We investigate the fate of a many-body-localized phase in a periodically driven non-Hermitian
chain of hard-core bosons with directed hopping subject to a binary onsite potential in the form of a deterministic quasiperiodic Fibonacci sequence. While Floquet heating generally results in the thermalization of the many-body-localized system with increasing driving period, our initially many-body-localized phase displays a sequence of localization-delocalization transitions, leaving isolated extended islands between many-body-localized phases. Non-Hermiticity makes these extended islands strongly boundary sensitive, and gives rise to a Floquet many-body skin effect under open boundary conditions. We show that the re-entrant many-body localization~(MBL) is independent of our choice of potential by reproducing this phenomenology when instead considering a potential with a genuinely disordered random-dimer sequence. Furthermore, we find that the directed hopping plays a dual role: it stabilizes MBL in regimes that are only weakly localized in the Hermitian limit, while promoting delocalization and broadening extended regions elsewhere. Our work advances the fundamental understanding of non-Hermitian and nonequilibrium quantum matter by revealing how directed hopping and spatial correlations in the onsite landscape jointly govern MBL in periodically driven interacting systems.\\

\end{abstract}

\maketitle
\textit{Introduction.---}
Non-Hermitian (NH) systems have recently emerged as a powerful extension of quantum theory for describing open systems, where dissipation, decoherence, and gain-and-loss are intrinsic~\cite{RevModPhys.93.015005,Ashida02072020,PhysRevX.9.041015,Banerjee_2023}. Their nonunitary dynamics produce phenomena without Hermitian counterpart, most prominently the NH skin effect, in which an extensive number of eigenmodes accumulate at the boundary and the conventional bulk–boundary correspondence breaks down~\cite{PhysRevLett.121.086803,PhysRevB.97.121401,PhysRevB.99.201103,PhysRevLett.124.086801,PhysRevLett.121.026808,PhysRevLett.123.066404,Gohsrich_2025}. Non-Hermiticity also reshapes disorder-induced localization~\cite{PhysRevLett.77.570,PhysRevE.59.6433,PhysRevB.58.8384,PhysRevE.63.036222,PhysRevB.103.054203,PhysRevLett.126.106803}. In interacting systems it gives rise to many-body skin effects with pronounced boundary sensitivity~\cite{PhysRevLett.133.136502,PhysRevB.111.035144,PhysRevLett.133.136503,PhysRevLett.133.076502,PhysRevB.108.165420} and unconventional NH many-body localization (MBL) transitions~\cite{PhysRevB.102.064206,PhysRevA.103.033325,PhysRevB.107.L220205,PhysRevLett.134.180405,PhysRevLett.123.090603,PhysRevB.106.064208}. Like their Hermitian counterparts~\cite{PhysRevB.87.134202,PhysRevLett.119.260401,PhysRevB.106.144201,PhysRevB.106.L220201}, these transitions are sensitive to whether the onsite landscape is quasiperiodic or spatially correlated.

\begin{figure}[h!]
\centering
     \includegraphics[width=0.479\textwidth]{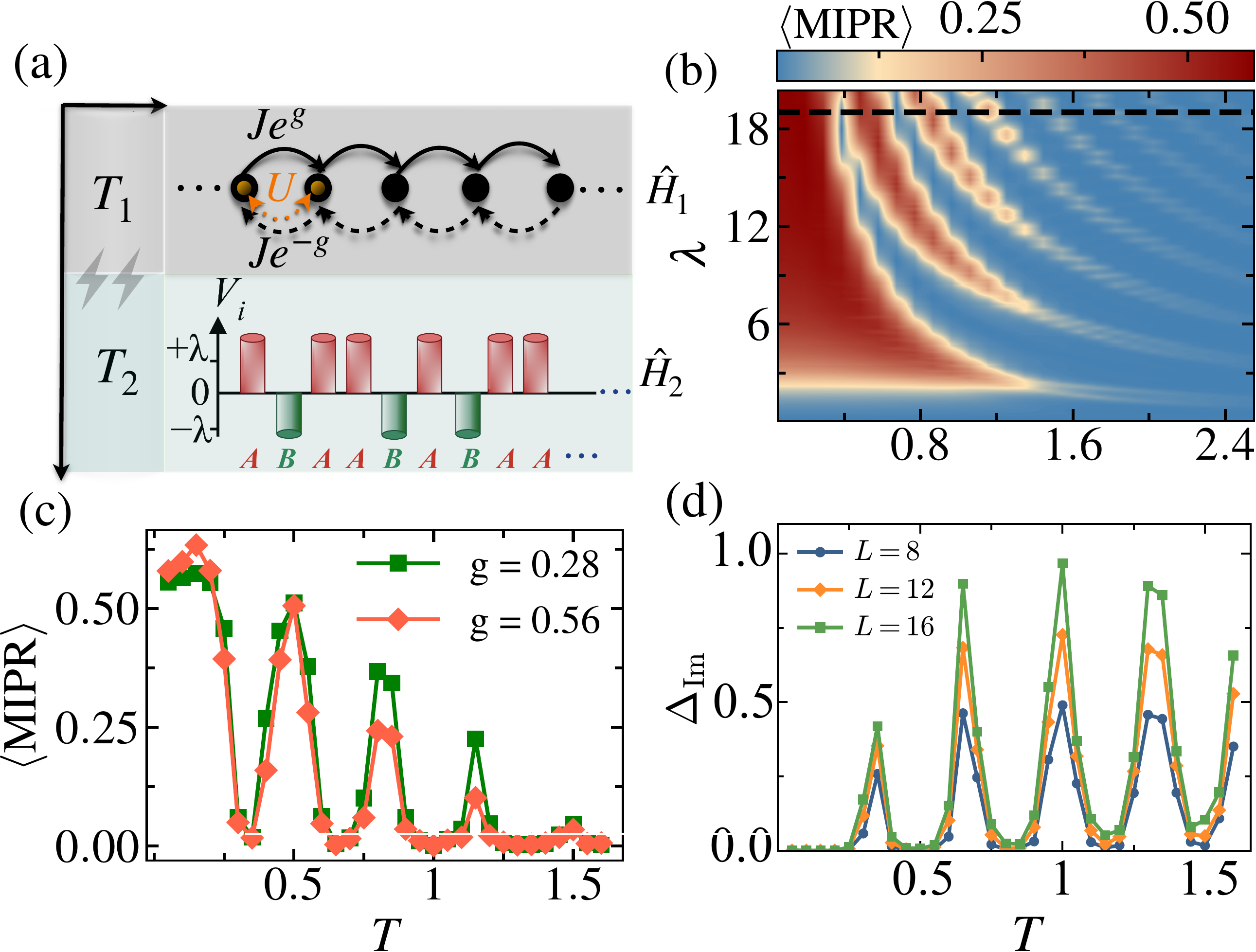}  
\caption{Re-entrant phenomena via spectral signatures in Fibonacci-dimer landscape. (a) Schematic depiction of the periodically driven NH many-body system. Driving with periods $T_1$ and $T_2$ alternates the system between Hamiltonians $  \hat{H_1}$, cf. Eq.~\eqref{eq:H1}, and $\hat{H_2}$, cf. Eq.~\eqref{potential}, respectively. (b) Phase diagram of $\langle \mathrm{MIPR} \rangle$ as a function of $T$ and $\lambda$ for $L=14$ and NH parameter $g=0.28$.
(c) Cut at $\lambda=19$ (black dashed line in (b)): $\langle \mathrm{MIPR} \rangle$ versus $T$ for $g=0.28$ (green squares) and $g=0.56$ (orange diamonds). 
(d) At the same cut, maximum imaginary energy ($\Delta_{\mathrm{Im}}$) versus $T$ for $g=0.28$ and system sizes $L=8$ (blue circles), $L=12$ (orange diamonds), and $L=16$ (green squares). We set $J=1$ and $U=2$.
}
\label{fig:phase-dig}
\end{figure}

Periodic driving is the most stringent test of MBL. In generic isolated, interacting systems, periodic driving induces Floquet heating, driving the system toward an infinite-temperature thermal state~\cite{peng2021floquet,rudner2020band,PhysRevLett.115.256803,PhysRevB.93.155132,DALESSIO201319}. In Hermitian systems, MBL survives only below a critical driving period that grows with disorder strength; beyond it, the system undergoes an MBL-to-ergodic transition~\cite{PhysRevLett.114.140401,PhysRevLett.115.030402,PhysRevE.90.012110,bordia2017periodically}. Whether this picture survives nonunitary dynamics is unknown. On top of that, non-Hermiticity adds ingredients absent in the Hermitian case. Complex quasienergies let a few amplified states dominate the long-time dynamics, and the skin effect drives directional accumulation at the boundary. Either could suppress or accelerate Floquet heating. Correlated disorder sharpens these deliberations further: random-dimer potentials host resonant delocalized states~\cite{PhysRevLett.65.88}, and display nontrivial localization under driving already at the single-particle level~\cite{PhysRevB.102.134212}. This raises two central questions: does non-Hermiticity stabilize or destroy MBL under periodic driving, and how do spatial correlations in the onsite potential influence this competition?

In this work, we address these questions by considering a periodically driven one-dimensional lattice of interacting hard-core bosons with directed hopping and a correlated binary onsite potential, initially arranged according to a deterministic Fibonacci sequence (see Fig.~\ref{fig:phase-dig}(a)). Using a symmetric quench protocol, we show that the system undergoes MBL-to-extended-to-MBL transitions, cf. the red (MBL) and blue (extended) regions in Fig.~\ref{fig:phase-dig}(b), where the critical driving period of a transition depends on the Fibonacci potential strength ($\lambda$) as well as the driving period ($T$), as illustrated by Fig.~\ref{fig:phase-dig}(c), before eventually delocalizing at a large driving period.
Concurrently, these transitions correspond to a shift from predominantly real quasienergy spectra in MBL phases to complex spectra in extended phases, cf. Fig.~\ref{fig:phase-dig}(d). The latter in turn gives rise to a robust Floquet many-body skin effect (FMBSE) under open boundaries, underpinning the boundary sensitivity in these extended regimes (islands), absent in Hermitian systems. To establish that this behavior is not specific to Fibonacci quasiperiodicity, we further rearrange the same binary onsite values into a genuinely disordered random-dimer sequence and observe a similar series of multiple re-entrant MBL transitions. Remarkably, directed hopping plays a dual role: it strengthens localization and stabilizes the MBL phase in certain regions, while favoring delocalization and broadening the extended region in others.

\textit{Model.---}
We consider a two-step periodic quench alternating between the Hamiltonians $\hat H_1$ and $\hat H_2$ for durations $T_1$ and $T_2$, respectively, as illustrated in Fig.~\ref{fig:phase-dig}(a). The first step is governed by an interacting Hatano--Nelson chain~\cite{PhysRevLett.77.570,PhysRevB.56.8651},
\begin{equation}
\hat H_1=
\sum_i\left[
-J\left(
e^{g}\hat b_{i+1}^{\dagger}\hat b_i
+e^{-g}\hat b_i^{\dagger}\hat b_{i+1}
\right)
+U\hat n_i\hat n_{i+1}
\right],
\label{eq:H1}
\end{equation}
where $\hat b_i$ ($\hat b_i^\dagger$) annihilates (creates) a hard-core boson at site $i$, $\hat n_i=\hat b_i^\dagger\hat b_i$, $J$ is the hopping amplitude, and $U$ is the nearest-neighbor interaction strength. The parameter $g$ controls the hopping asymmetry and renders the model non-Hermitian for $g\neq0$.

The second step introduces a correlated binary onsite potential, 
\begin{equation}
\hat H_2=\sum_i V_{\sigma_i}(\lambda)\hat n_i,
\label{potential}
\end{equation}
where $\sigma_i\in \{A,B\}$, $V_A(\lambda)$ and $V_B(\lambda)$ are the two onsite energies, and the sequence ${\sigma_i}$ determines their spatial arrangement. We choose $V_A=\lambda$ and $V_B=-\lambda$. For the Fibonacci potential, the sequence is generated deterministically through the substitution rules $A\rightarrow AB$ and $B\rightarrow A,$ which produces a quasiperiodic binary landscape~\cite{PhysRevResearch.3.033257,PhysRevB.104.014202,RevModPhys.93.045001}.

The Floquet operator over one driving period $T=T_1+T_2$ is
$
\hat F(T)=
e^{-i\hat H_1T_1}
e^{-i\hat H_2T_2}.
$
We numerically study the system under periodic boundary conditions (PBCs) at half filling, i.e., within the fixed-number subspace with total particle number $N=L/2$, where $L$ is the system size. We use a symmetric quench $(T_1 = T_2 = T/2)$ protocol. Interestingly, the stroboscopic evolution under $\hat{F}$ maps to a kicked Harper model with directed hopping in the non-interacting limit~\cite{PhysRevLett.65.3076,PhysRevLett.69.3302,PhysRevLett.87.066601,PhysRevB.106.054307}.

\textit{Static limit.---} 
Before turning to the driven problem, we
recall what is known in the static limit. In our protocol, this limit
is recovered at high driving frequency: for $T=T_1+T_2\to0$, the Floquet
operator reduces to $\hat U_F= e^{-i\hat H_{\mathrm{eff}}T}$ with
$\hat H_{\mathrm{eff}} \approx (T_1\hat H_1+T_2\hat H_2)/T$. This static Fibonacci chain presents an intriguing localization problem, and its behavior depends
sensitively on which ingredients are present. 
In the noninteracting Hermitian limit, $g = 0 = U$, the single-particle states remain critical for any finite potential strength $\lambda$, neither fully delocalized nor exponentially localized~\cite{MONTHUS}.
In the interacting case, $U \neq 0$, the Fibonacci chain undergoes an ergodic-to-MBL transition with increasing potential strength~\cite{PhysRevResearch.1.032039,10.21468/SciPostPhys.6.4.050}, which is surprising as interactions generally introduce additional scattering channels that would naively favor delocalization. If we instead consider directed hopping $g \neq 0$ but no interactions $U=0$, even an arbitrarily small hopping asymmetry completely delocalizes the noninteracting Fibonacci chain under PBCs, irrespective of the potential strength~\cite{PhysRevB.104.014202}. Whether interactions can sustain MBL when the underlying single-particle states are already delocalized by directed hopping is therefore far from obvious. Periodic driving makes this question even more compelling. While it generally promotes heating and delocalization in interacting Hermitian systems, here it acts alongside directed hopping, which already favors delocalization in the noninteracting limit. Both effects might therefore be expected to work against localization, making the fate of MBL under their combined influence particularly intriguing.

\textit{Phase diagram.---} In the following, we discuss the combined effect of interactions, directed hopping, a Fibonacci potential and driving, which results in a remarkable phase diagram depicted in Fig.~\ref{fig:phase-dig}(b). The phase diagram depicts the MBL and the extended regions marked by red and blue shades, respectively, which is obtained by plotting $\langle \text{MIPR} \rangle$ as a function of $T$ and $\lambda$, where
\begin{equation}
\text{MIPR} = \frac{1}{1-\nu} \left[ \frac{1}{\nu L} \sum_{i=1}^L \bar{n}_i^2 - \nu \right]
\label{eq:mipr}
\end{equation}
is the many-body inverse participation ratio (MIPR)~\cite{PhysRevLett.128.146601,PhysRevB.107.035129}. In Eq.~\eqref{eq:mipr}, \(\bar{n}_i = \langle \psi_\alpha | \hat{n}_i | \psi_\alpha \rangle\) is the local particle density, and \(\nu = 1/2\) is the filling factor. Here, $\langle \cdot\rangle$ stands for average over all the right eigenstates \(|\psi_\alpha\rangle\)~\cite{PhysRevLett.134.180405} with eigenstate index $\alpha$, which we obtain via exact diagonalization of the Floquet operator $\hat{F}$. We consider PBCs here to isolate bulk MBL from the FMBSE, which is only visible under open boundary conditions~(OBCs).

\begin{figure}[t]
\centering
\includegraphics[width=0.4896\textwidth]{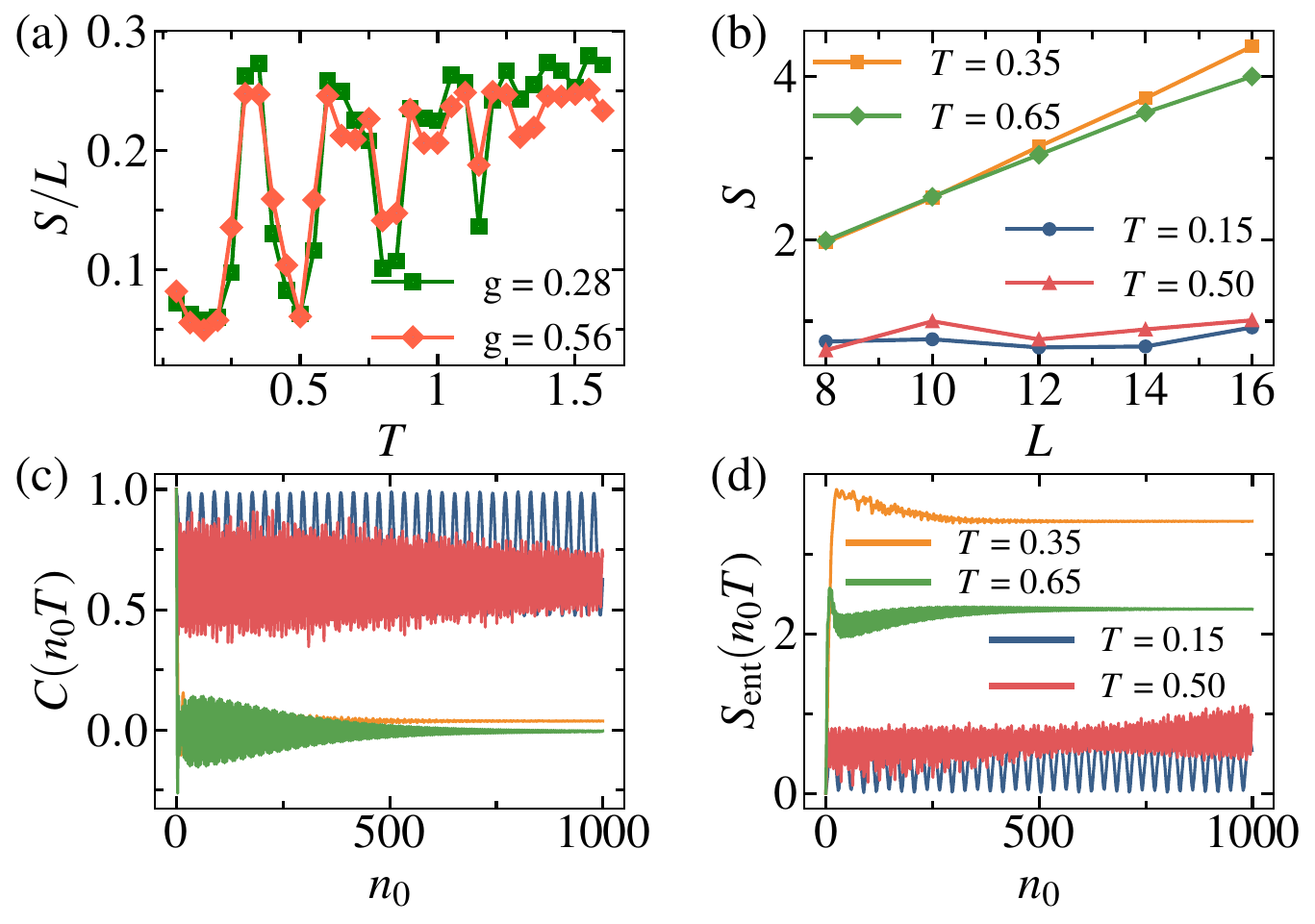}
    \caption{ Entanglement signatures and dynamical characterization in Fibonacci case. (a)~The normalized half-chain entanglement entropy \( S/L \) as a function of \( T \) for $g=0.28$ (green squares) and $g=0.56$ (orange diamonds) with $L = 16$.  (b)~Finite-size scaling of \( S \) at $g=0.28$ corresponding to $T=0.15$ (blue circles), $T=0.35$ (orange squares), $T=0.50$ (red triangles) and $T=0.65$ (green diamonds). (c)~The autocorrelation function \( C(n_0T) \) at $g=0.28$ shown for various driving periods: $T=0.15$ (blue), $T=0.35$ (orange), $T=0.50$ (red) and $T=0.65$ (green) for a system of size $L=24$ as a function of the number of drive cycles $n_0$. (d) Entanglement entropy \( S_{\text{ent}}(n_0 T) \) at $g=0.28$, for different driving periods as shown in (c). For all plots, we set $\lambda=19$, $J=1$ and $U=2$.
}
\label{fig:cut}
\end{figure}
The MIPR quantifies the particle distribution in real space, where $\langle\text{MIPR}\rangle \to 0$ for the extended phase and $\langle\text{MIPR}\rangle \to 1$ for the localized phase.
From the phase diagram in Fig.~\ref{fig:phase-dig}(b), it can be seen that the system undergoes an extended (blue region)
to MBL (red) phase transition in the low-driving limit ($T\to0$) as $\lambda$ increases. Upon increasing the driving period, the system experiences a transition from the MBL phase to an extended phase and subsequently to an MBL phase for large $\lambda$, constituting a re-entrant MBL transition in a driven system. This re-entrant MBL turns into multiple MBL transitions in the limit of large $\lambda$. Interestingly, the critical driving period of a transition tends to decrease with increase in the Fibonacci potential strength. This MBL-extended-MBL trend continues to appear with further increase in the driving period before the system completely delocalizes at large drive period.

In order to visualize this behaviour, we plot $\langle\text{MIPR}\rangle$ (green squares) as a function of $T$ in Fig.~\ref{fig:phase-dig}(c) for a cut (the black dashed line) through the phase diagram in Fig.~\ref{fig:phase-dig}(b) at $\lambda=19$. 
Initially in the limit of $T \to 0 $, the $\langle\text{MIPR}\rangle$ is at a maximum, which is typical for a strong MBL phase. With increase in $T$,  $\langle\text{MIPR}\rangle$ reaches a minimum near $ T \approx 0.354 $, indicating a transition of the system to the extended phase. However, with further increase in $T$,  $\langle\text{MIPR}\rangle$ increases again, signaling a re-entrant MBL phase transition. Subsequently, $\langle\text{MIPR}\rangle $ exhibits sharp nonmonotonic oscillations between low and high values, indicating multiple MBL transitions before eventually stabilizing in an extended phase. We note that while in the weak $T$ limit, $\langle\text{MIPR}\rangle$ remains robust against the parameter $g$, in the limit of large $T$, it decreases with increase in $g$, which can be seen from Fig.~\ref{fig:phase-dig}(c) (orange diamonds).

\textit{Real-complex spectral transitions.---} These re-entrant MBL transitions coincide with real–to-complex spectral transitions~\cite{PhysRevLett.123.090603,weidemann2022topological} of the Floquet quasienergies $\varepsilon$. In MBL regimes the quasienergies are predominantly real, whereas in extended regions they acquire finite imaginary parts due to directed hopping. 
To illustrate the correspondence between these simultaneous phase transitions, we plot $\Delta_{\mathrm{Im}} = \max_\alpha \big|\operatorname{Im}[\varepsilon_\alpha]\big|$
as a function of $T$ for a cut through $\lambda = 19$ (black dashed line in Fig.~\ref{fig:phase-dig}(b)) in Fig.~\ref{fig:phase-dig}(d). We see that $\Delta_{\mathrm{Im}}$ in the range of $T$, where the extended islands ($\langle \mathrm{MIPR} \rangle \approx 0$) exist, increases with system size, while in the MBL regimes (finite $\langle \mathrm{MIPR} \rangle$), it remains near zero. We observe the same pattern across varying system sizes.


\textit{Entanglement entropy transitions.---} This behavior is further corroborated by the half-chain entanglement entropy $S_\alpha = -\mathrm{Tr}_{L/2}\bigl[\rho_\alpha \log \rho_\alpha \bigr]$, where 
$\rho_\alpha = \mathrm{Tr}_{L/2}\bigl[\,|\psi_\alpha\rangle \langle \psi_\alpha|\,\bigr].$ We average $S_\alpha$ over those right eigenstates, whose quasienergies lie within $\pm 10\%$ from the middle of the real part of the spectrum, as well as over numerous realizations different Fibonacci configurations. 
The normalized entropy  $S  / L$ plotted in Fig.~\ref{fig:cut}(a) for $\lambda=19$, starts with a low value in the MBL phase, rises sharply with $T$ indicating an extended regime, and then oscillates non-monotonously with driving period. These entropy fluctuations suggest multiple MBL transitions before stabilizing in an extended phase at large $T$. Moreover, the system-size scaling of $S$ shown in Fig.~\ref{fig:cut}(b), reveals area-law saturation in the MBL phases (e.g., at \( T = 0.15 \) and \( T = 0.5 \)), and volume-law growth~\cite{PhysRevLett.123.090603,PhysRevB.108.184205} in extended phases (e.g., at \( T = 0.35 \)  and \( T = 0.65 \)).

\textit{Physical picture.---} The Fibonacci potential landscape, through its interplay with the periodic driving protocol, offers an intuitive framework for interpreting our results. Although directed hopping completely delocalizes the noninteracting Fibonacci chain under PBCs, interactions can support MBL in the presence of a sufficiently strong potential. Consequently, as $\lambda$ increases at short periods $T$, the system undergoes an MBL transition, because the driving frequency $\omega \sim 1/T$ is too large for efficient energy absorption. At large $T$, beyond a critical threshold, the slow driving acts as a sequence of nonadiabatic quenches that repeatedly dephase the system and can promote delocalization by reducing the effectiveness of disorder~\cite{PhysRevLett.115.030402}. At intermediate $T$, we observe hyperbolic-shaped extended regions that originate from an intriguing mechanism: the onsite kick of Floquet driving is diagonal in the occupation basis and factorizes into site-dependent local onsite phases. At particular combinations of the potential strength and kick duration $T_2$, these phases become spatially uniform, such that the onsite evolution $e^{-i\hat H_2T_2}$ reduces to a global phase $e^{-i\theta\hat N}$ within a fixed-particle-number sector. This effectively removes the spatial structure of the binary landscape from the Floquet evolution and suppresses localization during the corresponding driving cycle. Since this phase-matching mechanism relies only on the binary character of the onsite potential, rather than its spatial arrangement, it is expected to generalize beyond the case of Fibonacci quasiperiodicity. This observation naturally raises the question of whether the same re-entrant behavior persists for genuinely disordered but spatially correlated binary potentials.

\begin{figure}[t]
\centering
   \includegraphics[height=0.714\columnwidth]{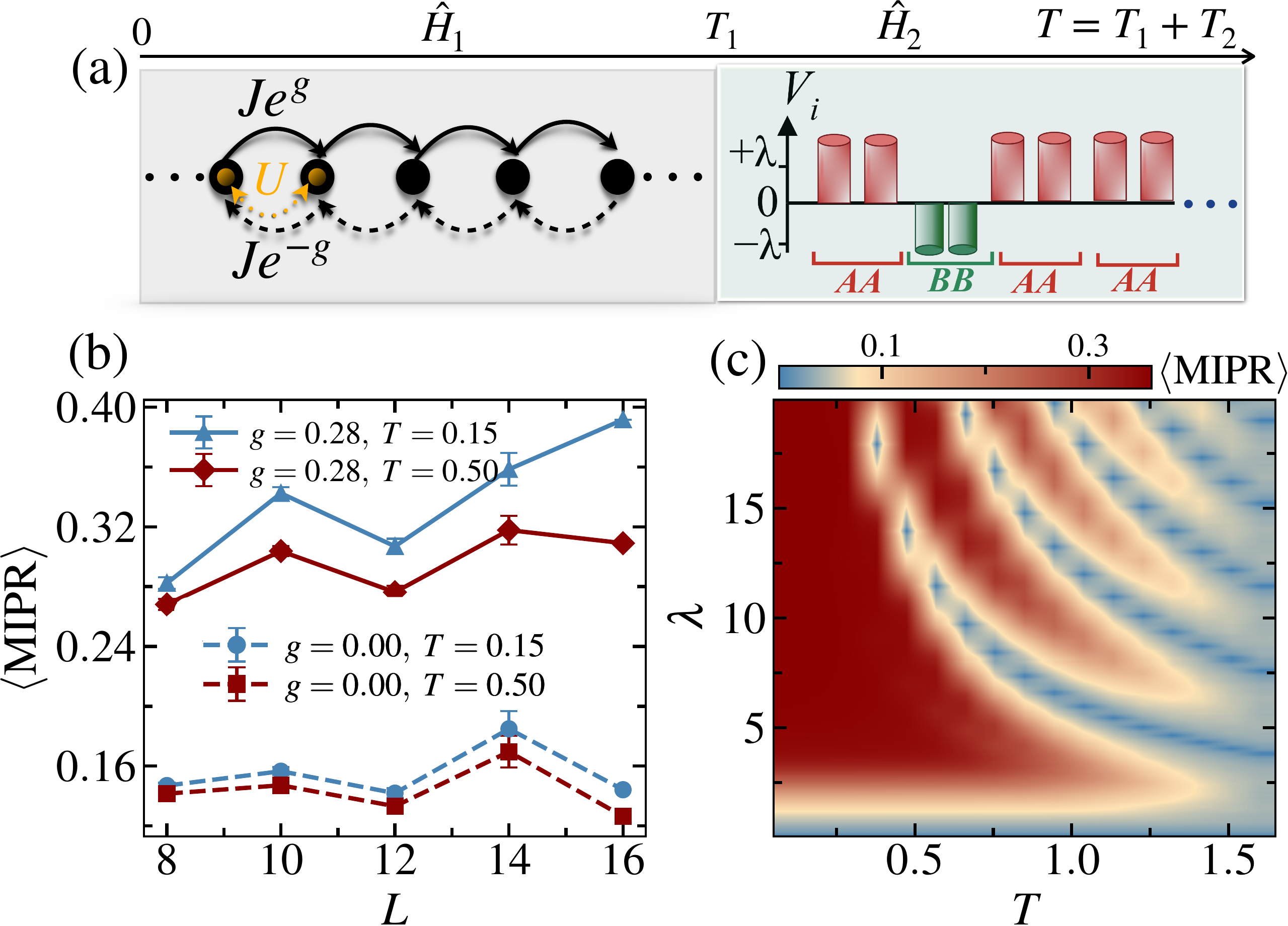}
    \caption{Random-dimer scenario. (a) Same driving protocol as shown in Fig.~\ref{fig:phase-dig}(a) but with a random dimer landscape in $\hat{H}_2$. (b) Finite-size scaling of $\langle\mathrm{MIPR}\rangle$ for $g=0$, $T=0.15$ (blue circles, dashed), $g=0$, $T=0.50$ (red squares, dashed), $g=0.28$, $T=0.15$ (blue triangles, solid), and $g=0.28$, $T=0.50$ (red diamonds, solid), showing enhanced localization due to directed hopping. (c) MIPR phase diagram for $g=0.28$, revealing multiple re-entrant MBL--extended--MBL transitions. Other parameters are $ J=1$, $\lambda = 19$, and $U=2$. }
\label{fig:RMDD}
\end{figure}

\paragraph*{Random-dimer case.---}

One example of such a spatially correlated binary potential is a random-dimer landscape~\cite{PhysRevLett.65.88,PhysRevA.106.013305,PhysRevA.81.041604}. To implement this in our model, we replace the deterministic sequence $\sigma_i$ in Eq.~\eqref{potential} by a random-dimer arrangement while retaining the same binary onsite energies. Specifically, the sequence ${\sigma_i}$ is constructed as $\sigma_{2\ell-1}=\sigma_{2\ell}\equiv\xi_\ell$, with $\xi_\ell\in\{A,B\}$, where each $\xi_\ell$ is chosen independently with equal probability~\cite{PhysRevLett.65.88,PhysRevB.102.134212}, see Fig.~\ref{fig:RMDD}(a) for a schematic depiction. The potential landscape generated in this way shares the same binary onsite values as in the Fibonacci case but differs fundamentally in its spatial correlations: the Fibonacci sequence is deterministic and quasiperiodic, whereas the random-dimer sequence is genuinely disordered. Figure~\ref{fig:RMDD}(b) shows that, in the Hermitian limit $g=0$, the $\langle \mathrm{MIPR} \rangle$ at $T=0.15$ and $T=0.50$ remains relatively small, $\langle\mathrm{MIPR}]\rangle\sim0.15$, indicating only weak localization rather than a stable MBL phase. Remarkably, directed hopping qualitatively reshapes these regimes: for $g=0.28$, the MIPR increases to approximately $0.4$ and remains robust with increasing system size, demonstrating a substantial enhancement of localization and stabilization of the MBL phase. Consistently, the phase diagram in Fig.~\ref{fig:RMDD}(c) exhibits multiple re-entrant MBL--extended--MBL transitions. Directed hopping therefore plays a dual, parameter-dependent role in the random-dimer system: it stabilizes emergent MBL regions in some regimes, while promoting delocalization and broadening the extended region at larger driving periods, analogous to the behavior found for the Fibonacci potential in Fig.~\ref{fig:phase-dig}(c).

\begin{figure*}[t]
\centering
    \includegraphics[width=2.0\columnwidth]{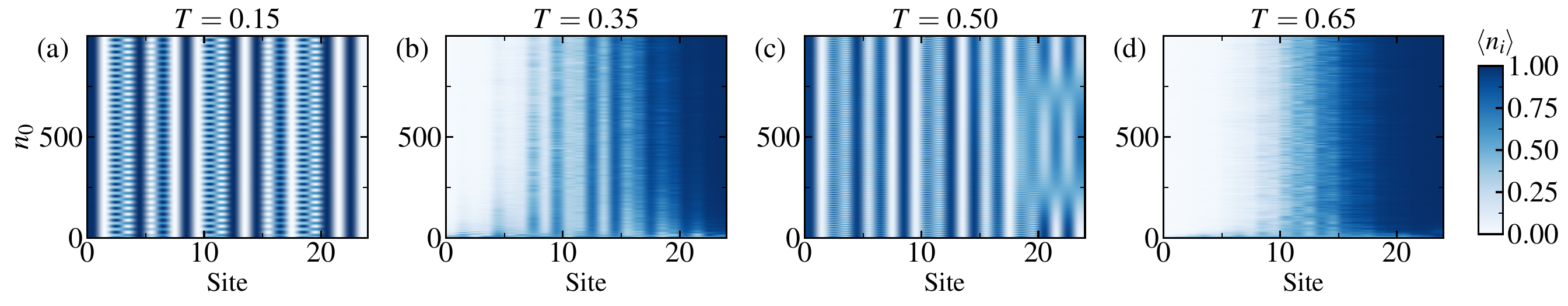}
    \caption{Onsite density $\langle n_i\rangle$  as a function of lattice site ($i$) and driving cycle $n_0$ for a system of size \( L = 24 \) under OBCs with Fibonacci landscape. Panels (a) and (c) correspond to \( T = 0.15 \) and \( T = 0.50 \), respectively, where the density profile preserves the initial CDW pattern over time, indicating re-entrant MBL. In contrast, panels (b) and (d) for \( T = 0.35 \) and \( T = 0.65 \), respectively, reveal strong, boundary accumulation characteristic of Floquet many-body skin modes, arising from NH wavepacket dynamics. We consider $g = 0.28$, $\lambda = 19$, $J=1$, and $U=2$.}
\label{fig:skin-effect}
\end{figure*}

\textit{Dynamics.---} Having established that the re-entrant structure persists beyond Fibonacci quasiperiodicity, we now return to the Fibonacci realization and examine experimentally accessible dynamical observables~\cite{bordia2017periodically} to assess the stability of its re-entrant MBL phases under PBCs. We study nonequilibrium evolution within the framework of Lindbladian dynamics, focusing on trajectories, post-selected with no quantum jumps~\cite{Daley04032014}.
Starting from a charge-density wave (CDW) as an initial state \( |\psi_0\rangle = |0101\cdots\rangle \), the stroboscopic evolution is computed using the non-unitary Floquet operator \( \hat{F} \), with \( |\psi(n_0T)\rangle = \hat{F}(n_0T)|\psi_0\rangle \), normalized at each step as \( |\widetilde{\psi}(n_0T)\rangle = |\psi(n_0T)\rangle / \||\psi(n_0T)\rangle\| \)~\cite{PhysRevLett.123.090603,PhysRevB.106.064208,mak2024statics,PhysRevB.109.L140201,PhysRevB.108.184205,PhysRevB.101.184201}. Here, $n_0$ denotes the number of drive cycles. For $L=24$, we simulate the dynamics using the Krylov subspace method with Arnoldi iteration~\cite{bhattacharya2023krylov,bhattacharya2022operator, PhysRevB.111.064203, NANDY20251, PhysRevB.105.024303,PhysRevB.108.214308}. We focus on two observables: the temporal autocorrelation function \( C(n_0T) = \frac{1}{L} \sum_{i} C_i(n_0 T) \), where \( C_i(n_0 T) = (2\langle n_i(n_0T)\rangle - 1)(2\langle n_i(0)\rangle - 1) \) and $
\langle \hat n_i(n_0T)\rangle
=\langle \widetilde{\psi}(n_0T)|\hat n_i|\widetilde{\psi}(n_0T)\rangle$,
and the half-chain entanglement entropy \( S_{\text{ent}}(n_0 T) = -\mathrm{Tr}[\rho_A(n_0 T) \ln \rho_A(n_0 T)] \)~\cite{PhysRevLett.134.180405, PhysRevB.105.024303,PhysRevB.108.214308,Felski_2024}, with \( \rho_A(n_0 T) \) being the reduced density matrix of the subsystem $A$. In Fig.~\ref{fig:cut}(c) and (d), we plot $C(n_0 T)$ and $S_{\text{ent}}(n_0 T)$, respectively, for different values of $T$ chosen from the MBL and extended regions.
In the ergodic regimes (e.g., \( T = 0.35, 0.65 \)), \( C(n_0T) \) decays rapidly to zero, indicating the loss of CDW memory. For $T = 0.35$, the entanglement entropy $S_{\text{ent}}(n_0 T)$ grows and saturates at a high, $L$-dependent value, while at $T = 0.65$, it exhibits non-monotonic behavior and saturates at a relatively lower value, due to enhanced $\Delta_\text{Im}$ and nonunitary dynamics~\cite{PhysRevB.105.024303}. However, vanishing \( C(n_0T) \) in both $T$ provides clear evidence of ergodicity~\cite{PhysRevLett.134.180405}. 
In the MBL phases (e.g., for \( T = 0.15, 0.5 \)), \( C(n_0 T) \) remains close to unity, exhibiting persistent oscillations that reflect the fingerprint of the underlying Fibonacci potential and CDW initial state~\cite{PhysRevB.100.104203,PhysRevResearch.3.033257}. Additionally, \( S_{\text{ent}}(n_0 T) \) for the MBL phase exhibits slow growth~\cite{PhysRevLett.123.090603,PhysRevB.108.214308,PhysRevB.105.024303,PhysRevB.101.184201} with oscillations, consistent with area-law scaling. The absence of Griffiths effects~\cite{PhysRevB.100.104203} in quasiperiodic systems ensures finite  saturation of $C(n_0 T)$ in the MBL phase without any substantial late-time decay (see Fig.~\ref{fig:cut}(c)).

\textit{Competing MBL and FMBSE.---} Systems with directed hopping, both at the single-particle and many-body interacting regime, exhibit boundary-sensitive dynamical behavior under OBCs due to the NH skin effect~\cite{PhysRevLett.121.086803,PhysRevB.97.121401,Gohsrich_2025,PhysRevA.104.022215,PhysRevB.107.L220205,PhysRevLett.133.136503,shen2022non,PhysRevLett.133.136502,PhysRevB.111.035144}. To probe the boundary sensitivity of the MBL and the extended phases, we study wavepacket dynamics under OBCs, starting from a CDW initial state. We monitor the evolution of onsite densities $\langle n_i\rangle$ over successive drive cycles $n_0$, as shown in Fig.~\ref{fig:skin-effect}. In the MBL phases (e.g., for $T=0.15$ and $T=0.5$), the CDW order remains largely preserved, with no pronounced particle accumulation at the boundary (see Figs. \ref{fig:skin-effect}(a) and (c)), indicating that the localized regimes identified under PBCs remain robust upon opening the boundaries, consistent with their predominantly real quasienergies. This robustness provides a basis for using the PBC phase diagram to guide our investigation of boundary-sensitive dynamics in the extended islands. In contrast, within the extended islands, where complex quasienergies dominate under PBCs, we observe a robust manifestation of the FMBSE under OBCs. At representative driving periods \(T = 0.35\) and \(T = 0.65\) (Figs.~\ref{fig:skin-effect}(b) and (d)), the system rapidly loses memory of its initial CDW configuration, which is revealed by a clear asymmetry in the particle density accumulated at the right boundary. This behavior signals the emergence of the FMBSE, which is in stark contrast to localization-induced memory retention in the MBL phase. We note that, for $T=0.65$ (see Fig.~\ref{fig:skin-effect}(d)), a stronger FMBSE is observed as compared to that for $T=0.35$ (see Fig.~\ref{fig:skin-effect}(b)), accompanied by the enhanced $\Delta_\mathrm{Im}$ for the former case (compare Fig.~\ref{fig:cut}(d)). This analysis reveals that as the driving period is varied, the system alternates between MBL and displaying the FMBSE, underscoring the intricate boundary sensitivity and NH dynamics of driven quasiperiodic systems.

\textit{Conclusions.---} We uncover a conceptually new, drive-induced re-entrant MBL transition in a periodically driven NH many-body system with Fibonacci potentials, featuring multiple sharp extended–to-MBL transitions governed by the driving period. A similar re-entrant structure persists for a genuinely disordered random-dimer potential, demonstrating that the phenomenon extends beyond Fibonacci quasiperiodicity to correlated binary landscapes. In this case, directed hopping nontrivially stabilizes the MBL phase from a regime that remains only weakly localized in the Hermitian limit, thereby enabling re-entrant localization transitions.
 Experimentally measurable quantities from wavepacket dynamics reveal robust  FMBSE in these extended regimes, underscoring boundary sensitivity absent in Hermitian systems.
These findings facilitate applications in frequency-modulated quantum sensors~\cite{App1, App2, App3, App4, App5, App6} and quantum information processing~\cite{MBLapp1, MBLapp2}.
In addition, the rapid entanglement growth in extended Floquet regimes limits tensor network simulations, positioning quantum circuits as an attractive platform for realizing and benchmarking such dynamics. These unconventional non-equilibrium phases can be experimentally realized in ultracold atoms~\cite{PhysRevX.8.031079,PhysRevA.106.L061302,PhysRevLett.124.250402,zhao2025two,PhysRevLett.129.070401}, quantum circuits~\cite{zhang2025observation,zhang2022digital,Fauseweh2023quantumcomputing,kumar2025floquet,PhysRevLett.132.120402,eckstein2024large,yanay2020two,karamlou2024probing,PhysRevResearch.7.023032}, and superconducting platforms~\cite{yanay2020two}, opening avenues for driven NH many-body physics in quasiperiodic and higher-dimensional settings.

\paragraph{Acknowledgement.-} T.M. acknowledges support from Anusandhan National Research Foundation (ANRF), Government of India, through project No. ANRF/ARG/2025/007442/PS. A.B. and F.K.K. acknowledge funding from the Max Planck Society Lise Meitner Excellence Program 2.0.
We also acknowledge funding from the European Union’s ERC Starting Grant “NTopQuant” (101116680). Views and opinions expressed are however those of the authors only and do not necessarily reflect those of the European Union or the European Research Council (ERC). Neither the European Union nor the granting authority can be held responsible for them.

\bibliography{references}

\end{document}